\documentclass[twocolumn,showpacs,preprintnumbers,amsmath,amssymb]{revtex4}

\usepackage{graphicx}
\usepackage{pstricks,epsfig}
\usepackage{dcolumn}
\usepackage{bm,longtable,braket,ulem,pstricks}

\usepackage{color}

\begin{document}
	
\title{Many-Body electronic structure calculations of Eu doped ZnO}
\author{M. Lorke}
\email{mlorke@uni-bremen.de}
\author{T. Frauenheim}
\affiliation{Bremen Center for Computational Materials Science, University of Bremen, Am Fallturm 1, 28359 Bremen, Germany}
\author{A. L. da Rosa}
\affiliation{Universidade Federal de Minas Gerais, Dept. of Physics, Av. Ant\^onio Carlos, 6627, 31270-901 Belo Horizonte, MG, Brazil}
\date{\today}

\begin{abstract}
The formation energies and electronic structure of europium doped zinc
oxide has been determined using DFT and many-body $GW$ methods. In the
absence of intrisic defects we find that the europium-$f$ states are
located in the ZnO band gap with europium possessing a formal charge
of 2+. On the other hand, the presence of intrinsic defects in ZnO
allows  intraband $f-f$ transitions otherwise forbidden in atomic europium. This
result coorroborates with recently observed photoluminescence in the
visible red region\,\cite{Ronning2014}.
\end{abstract}


\maketitle

\section{Introduction}

Doping has been widely used to tailor the electronic, magnetic and
optical properties of semiconductors. Wide band-gap semiconductors
such as ZnO are attractive for ultraviolet light-emitting diodes,
lasers and high-power photonic applications. In ZnO, rar-earth
elements can be incorporated in the material and the long life times
of the excited states allow for an easy realization of population
inversion with promising applications in optoelectronic
applications\,\cite{Ishizumi:05,Peres:07,Pan:09,Ronning:10,Ronning2014}.
ZnO has a large band gap of 3.4 eV and a high thermal conductivity,
enabling new electroluminescent devices.

Channeling experiments \cite{Wahl:03,Wahl:04} indicate, that
rare-earth elements in ZnO are preferentially incorporated at cation
sites.  More recent photoluminescence (PL) and photoluminescence
excitation (PLE) investigations on ZnO nanoparticles corroborate this
finding \cite{Du:08,Ji:09}.  PL investigations of Eu-doped ZnO nanoneedles showed sharp emission lines
from Eu$^{3+}$, suggesting the emission arises from intra-4$f$
transitions, in addition to the ZnO interband emission
\cite{Ebisawa:08}. The f to f transitions are forbidden in the isolated atom. However the spherical symmetry is broken when the impurity is incorporated into the ZnO matrix. Besides Eu, optical
emission from rare-earth orbitals has been achieved in ZnO nanowires
implanted with erbium and ytterbium \cite{Ronning:10} and thin ZnO
films doped with erbium, samarium and europium
\cite{Petersen:10,Dai:11,Tsuji:12,Mezdrogina:12}.  The above
experimental investigations have suggested explanations for the
optical activation of rare-earths in ZnO \cite{Akazawa:13,Yang:14} by
means of substitutional hydrogen incorporation \cite{Akazawa:13} or
formation of defect complexes \cite{Du:08,Wang:11,Yang:14}.

Theoretical investigations using density functional theory (DFT) in
the generalized-gradient approximation (GGA) have performed for Eu
doped ZnO \cite{Assadi:11,Chen:13}, but were unable to ascertain the
origin of the experimentally observed emission in ZnO. The main
challenge here is the correct description of both ZnO band edges and
defect states. It is common understanding that the use of local
exchange-correlation functionals wrongly described the ZnO band gap,
which could lead to misleading conclusions on the location of the
impurity rare-earth $f$-states. It has been shown that the correct
description of the band gap is of paramount importance for the
understanding of impurity and defect states in
semiconductors\cite{Lany:08,Lany:09,Lany:10,Janotti:09,Sarsari:13}.
Besides, intrinsic defects may also play an important role, since
during ion implantation they are often introduced in ZnO.  Jiang et
al. \cite{Jiang:09} have shown that for many open $d$- or $f$-shell compounds
the GW approach can provide a consistent and accurate treatment for
both localized and itinerant states.

Using DFT calculations and the $GW$ technique, we have previously shown that a
complex containing a single oxygen interstitial defect and an europium
atom substituting a zinc atom is a probable candidate to explain the
observed emission in the red region of the spectrum in europium doped
ZnO nanowires,\cite{Ronning2014}. Note that GW calculations do not
include excitonic electron-hole interactions. 

In this work we show $GW_0$ calculations, as implemented in VASP, for Eu-doped
ZnO. Besides the work reported in Ref.\cite{Ronning2014} we have
considered the presence of the most common defects in ZnO, such as
oxygen and zinc vacancies and interstitials. The wave functions are
kept fixed to the GGA level, whereas the eigenvalues are updated in
the Green's function only. Here we show that this kind of defect has a
relatively low formation energy and suggest possible mechanisms for
the formation of such defect. Besides, we investigate other defects
and calculate their formation energies and electronic structure.

\section{Computational details}

\label{sec:theory}
In this work we have employed density functional theory (DFT)
\cite{Hohenberg:64,Kohn:65} and many-body $GW$ techniques
\cite{Hedin:65} to investigate the formation energies and
electronic structure of Eu doped ZnO. The projected augmented wave
method (PAW) \cite{Bloechel:94} has been used as implemented in the
Vienna Ab-initio Package (VASP) \cite{Kresse:99,Shishkin:07}.  A
3$\times$3$\times$2 supercell containing 72 atoms with a
(2$\times$2$\times$1) {\bf k}-point sampling and a cutoff of 500\,eV
is used to calculate all isolated intrinsic defects and complexes. 
For Eu$_2$O$_3$ we have used (2$\times$2$\times$2) k-points. For Eu metal we adopted a
bcc structure with a (6$\times$6$\times$6) {\bf k}-point mesh.

\subsection{Thermodynamic properties}
To verify the thermodynamic stability of the investigated defect complexes,
we follow the approach derived by van de Walle and Neugebauer\cite{JAP:2004}.
The formation energy of a neutral defect or impurity is defined as:

\begin{equation}
\label{formation}
E_{\rm f} = E_{\rm tot}^{\rm ZnO-defect} - E_{\rm tot}^{\rm ZnO-bulk} - \sum_i{n_i\mu_i} + {\rm q(E_v - E_F)},
\end{equation}

where E$_{\rm tot}$(\rm ZnO-defect) is the total energy of a defective
supercell and E$_{\rm tot}${\rm ZnO-bulk} is the total energy for the
supercell of pure ZnO. $n_i$ is the number of atoms of type $i$
(defects or impurities) that have been added to or removed from the
supercell and $\mu_i$ is the corresponding chemical potential of each
species, q is the charge of the defect. The Fermi level $E_F$ is referenced 
with respect to the valence band maximum $E_v$ of the host. 

The lower bound of the chemical potential corresponds to the total
absence of impurities/defects in the material. An upper bound on the
chemical potential is given by the energy of the elemental bulk phase
or other solubility-limiting phases. Formation of Eu clusters (metal) can be
avoided by imposing the europium chemical potential to obey to:

 \begin{equation}
\label{eq:eu}
 \mu_\text{Eu}\leqslant \mu_\text{Eu(Eu-bulk)}
 \end{equation}

where Eu-bulk is the total energy of europium bulk in a bcc structure.
Formation of europium precipitates are avoided by considering the bound
for Eu$_2$O$_3$, a very stable phase of europium:

 \begin{equation}
\label{eq:eu2o3}
 \mu_{\rm Eu}\leqslant \mu_{\rm Eu(Eu_2O_3)}
 \end{equation}

The oxygen and zinc chemical potential are not independent, but
related by $\mu_{\rm ZnO} = \mu_{\rm Zn} + \mu_O$, where $\mu_{\rm  Zn}$, $\mu_O$ and $\mu_{\rm ZnO}$ are the zinc, oxygen and zinc
oxide chemical potentials, respectively. $\Delta H^\text{ZnO}$ is the
formation enthalpy of ZnO.  Combining Eq.\ref{eq:eu} and
\ref{eq:eu2o3} and these conditions, we obtain the following
relations:

\begin{equation}
 \mu_O=\mu_{O_2-mol} + \lambda \Delta H^\text{ZnO}
 \end{equation}

The chemical potential of europium oxide can be writen as:

\begin{equation}
\mu_{\text{Eu}_2 O_3} =\Delta H^{\text{Eu}_2 O_3} + 2 \mu_\text{Eu-bulk} + 3 \mu_{O_2-mol}
\end{equation}

where $\Delta H^{\text{Eu}_2 O_3}$ is the enthalpy of formation of europium oxide and  $\mu_{\rm Eu-bulk}$ and 
$\mu_{\rm O_2-mol}$ are the europium bulk and oxygen molecule chemical potentials, respectively. This results in 

\begin{equation}
\mu_\text{Eu} \leqslant \frac{1}{2} \Delta H^{\text{Eu}_2 O_3} +  \mu_\text{Eu bulk} -\frac{3}{2}\lambda \Delta H^\text{ZnO}
\end{equation}

Here $\lambda$=0 for oxygen rich and  $\lambda$=1 for oxygen poor conditions.

\subsection{Electronic structure}

The many-body methodology that is underlying the $GW$ approximation
goes back to pioneering work by Kadanoff and Baym
\cite{Baym:61,Baym:62} and Hedin \cite{Hedin:65}.  
In $GW$-based approximations, the screening of the Coulomb interaction, that 
enters the non-local self-energy operator, is calculated microscopically. This  description of the screened potential $W$
is a main advantage of many-body methods over other approaches, like hybrid functional DFT, where screening is only included
phenomenologically.

The single shot G$_0$W$_0$ approximation starting from GGA calculations
often yields too small band gaps. It was suggested \cite{Holm:96,Shishkin:07} that 
partially or fully selfconsistent schemes,
in which either Green's function G ($GW_0$) or both the Green's function and the
dielectric matrix ($GW$) are updated can improve the agreement with experiments.
We should mention that the implementation of selfconsistent $GW$ or $GW_0$ in VASP has the shortcoming, that only calculations that maintain a quasi-particle picture
are possible, i.e. satellite peaks can not be accounted for. Even though this means that these are technically not ''full'' $GW_0$
calculations, we will for simplicity adopt the notation to call them $GW$ and $GW_0$ in the following.
For detailed information on the implementation we refer the reader to
Refs.~\cite{Shishkin:07,Rinke:05,Usuda,Sottile:12,Thygesen:2013,louie:2014}.  

\section{Results}
\subsection{Thermodynamic properties}

It is in principle possible to calculate GW total energies by using
the Galitskii-Migdal formula \cite{Fetter:03}. In Ref. \cite{Holm:99} it was found
that for the electron gas, the total energies can be well described using $GW$ schemes.
Alternative approaches to tackle the formation energy in connection with $GW$ methods have been proposed \cite{Rinke:09,Jain:11}. In VASP, the calculation of
formation energies at $GW$ level is not readily possible.  Therefore,
we have decided to calculate the formation energies at GGA level. This
can be justified taking into account that the GGA calculations are
used as starting point for more accurate calculations using the $GW$
method. We would like to mention that we have performed PBE0
calculations for some defects and we note that the location of
Eu-$f$-states are strongly dependent on the amount of HF included in
the functional.

For europium oxide, the formation enthalpy is calculated to be $\Delta
H^f(\rm Eu_2O_3)$=-14.43 eV compared with the experimental value of
-16.51 eV \cite{CRC}. For zinc oxide the formation enthalpy is $\Delta
H^f(\rm ZnO)$=-2.90 eV, which is in good agreement with other GGA
values\,\cite{Oba:11} and even GGA+U calculations \cite{Lany:10}. This
discrepancy is probably due to the wrong description of the oxygen
molecule, as has been discussed in Ref.\cite{Perdew:96}.  Using these
values, we can obtain the binding energy of the defect complexes,
which are defined as

\begin{equation}
{\rm E_{binding}= E_f^{complex} - \sum_i{E_f^{isolated}}}
\end{equation}

where ${\rm E_f^{complex}}$ is the formation energy of the defects and
${\rm E_f^{isolated}}$ is the formation energy of isolated defects
calculated according to Eq.\eqref{formation}. The calculated values are
shown in Table\,\ref{table:formation} and a negative value means that the complex is stable. Calculations for intrinsic
defects can reproduce well previous calculations reported using local
functionals \cite{Oba:10,Oba:11,Janotti:07}. Neutral oxygen vacancies have a
low formation energy under oxygen poor-conditions. On the other hand,
zinc vacancies can be formed under Zn-poor preparation conditions.
Oxygen and zinc interstitials, as expected, have high formation
energies, due to their size \cite{Janotti:07}.  
For Eu doped ZnO at cation site we obtain a formation energy of 2.42 eV under
O-rich conditions. Incorporation of Eu at interstitial positions is
highly unfavorable, due to the strain Eu causes in the ZnO, leading to
a strongly distorted lattice. Similar results have been reported in
Ref.~\cite{Ney:12}. Complex formation with zinc and oxygen
vacancies and zinc interstitials have a much higher formation
energy. This can be understood by considering size effects, which causes a large strain in the lattice.
As the defect complexes are formed in experiments under extreme non-equilibrium conditions 
(like, e.g.,ion-implantation) we find it more appropriate to report the binding energies of the 
defect complexes and hence their stability against dissociation.

As we can see the most stable neutral defect is under O-rich
conditions the Eu-O$_i$(o) complex. It is interesting to point out
that the formation of neutral oxygen defects at interstitial sites in
pure ZnO has a high formation energy under thermodynamic
equilibrium. However, as has been shown in Ref. \cite{Janotti:07} the
diffusion barrier for this kind of defect is relatively low (around
0.2 eV). Therefore, once this defect is formed under ion implantation
(non-equilibrium conditions), it can rapidly diffuse in the material
and form complexes with europium atoms. This may explain why the
complex Eu-O$_i$(o) is so stable in ZnO.

\begin{table}[ht!]
\begin{center}
\caption{\label{table:formation}Formation energies $E_f$ and binding energies E$_{\rm b}$ for neutral intrinsic defects and defect complexes in ZnO.}
\begin{tabular*}{0.5\textwidth}{@{\extracolsep{\fill}}lcccc}
\hline
\hline
Defect  & 	\multicolumn{2}{c}{${\rm E_{f}}$ (eV)} &   \multicolumn{2}{c}{\rm $E_{\rm b}$ (eV)} 	\\
\hline
       &	O-rich & O-poor & O-rich & O-poor \\ 
\hline
${\rm Eu_{Zn}}$ 	& 2.42   &  7.44 & &  \\
${\rm O_{i(split)}}$  & 2.58 & 5.48 & & 	\\
${\rm O_{i(oct)}}$  & 3.38 &  6.28 & & 	\\
${\rm V_{\rm O}}$ &   3.20 & 0.30 & &  \\ 
${\rm V_{\rm Zn}}$  &  1.37  &  4.27 & & 	\\
${\rm Zn_{int}}$ & 6.68 & 3.78 &  &  \\

${\rm Eu_{Zn}+O{i(split)}}$ & 4.52  & 14.67  & -0.48 & 1.75 \\
${\rm Eu_{Zn}+O{i(oct)}}$ & 4.4 & 14.55  & -1.40 & 0.83  \\
${\rm Eu_{Zn}+V_{O}}$ & 6.1  & 10.45 & 0.48 & 2.70 \\
${\rm Eu_{Zn}+V_{Zn}}$ & 10.9 & 13.80    & 7.11 & 2.09 \\
${\rm Eu_{Zn}+Zn_{int}}$ & 9.88 & 12.78  & 0.79 & 1.56  \\
\hline
\hline
\end{tabular*}
\end{center}
\end{table}

\begin{figure*}[ht!]
\includegraphics[width=\textwidth]{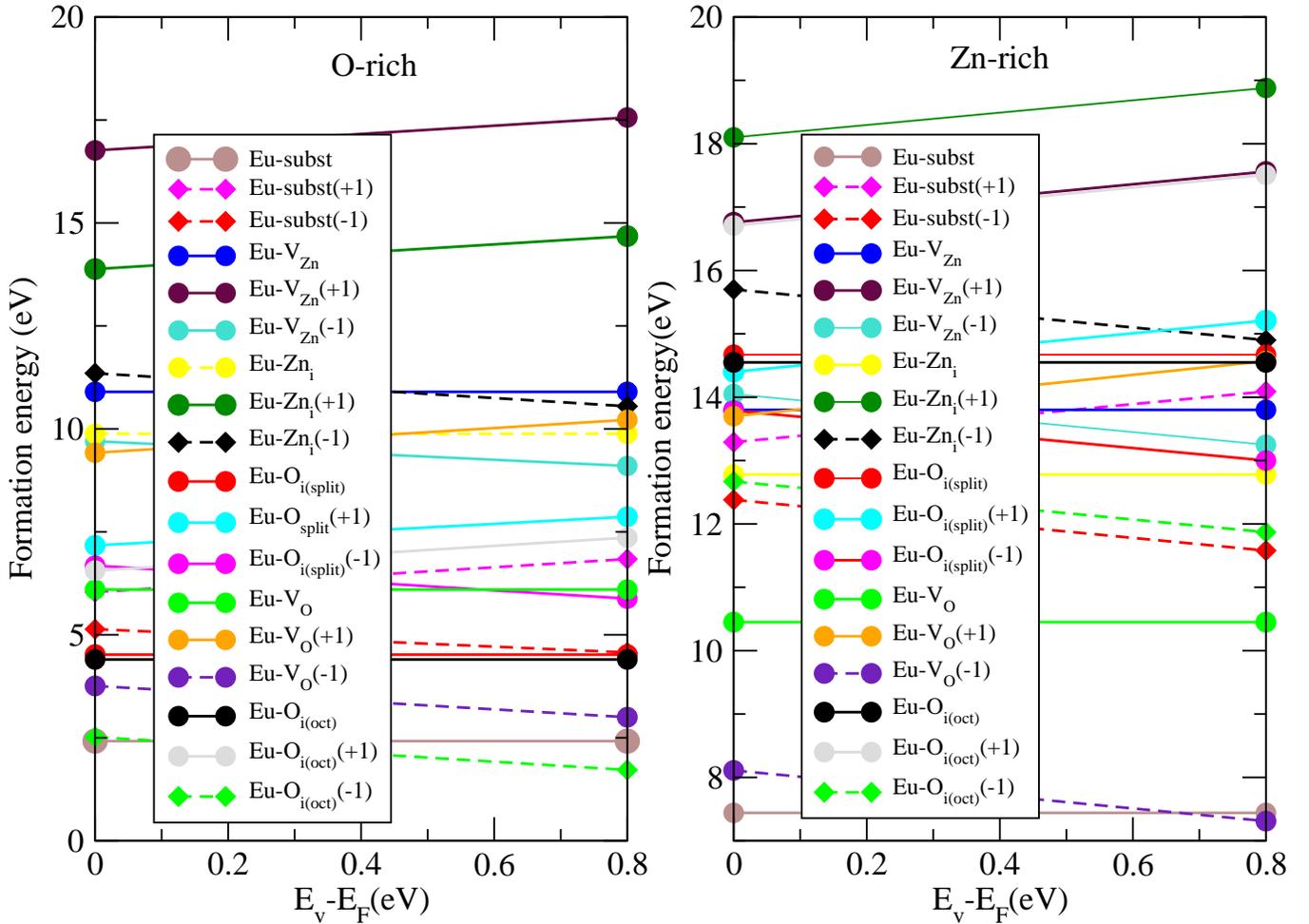}
\caption{\label{fig:formationall}(color online) Formation energies of several Eu-complexes as a function of the Fermi energy a) O-rich and b) Zn-rich conditions}. 
\end{figure*}

Now let us discuss the charged complexes. In  Fig.\ref{fig:formationall} we show the formation energies of several 
 Eu-complexes as a function of the Fermi energy at  O-rich and   Zn-rich conditions. The top of the valence band ${\bf E_v}$ for all defects  calculations was aligned with the top of the valence band of the
  host ZnO using the averaged electrostatic potential as described in
  Ref.\cite{JAP:2004}.  We see that the most stable structure at
  O-rich conditions is a complex of Eu-O$_i(oct)$ in the -1 charge
  state. Although this is the thermodynamically most stable defect, followed by Eu-subst and ${\rm Eu-V_O}(-1)$, 
we argue that the fact that the neutral Eu-O$_{i(oct)}$ 
correponds to the experimental f-f transition is very appealing. Obviouly we cannot rule out the co-existence of these other defects during the preparation conditions. Further calculations are needed to clarify whether these defects also have transitions in the experimentally observed region.  On the other hand at Zn-rich the defect Eu-subst is the most
  stable one for ${\rm E_F-E_v}$ between 0 and 0.6. For values lerger than that the defect Eu-V$_{\rm O}$
  in the -1 charge state is the most stable one. It is important to note that under Zn-rich conditions all defects
  have a very high formation energy, which is an indication that they
  are unlikely to form at these conditions.

\subsection{Electronic properties}

Because $GW$ calculations can be performed using
different approximations, the results for the theoretical band gap of
ZnO has been under debate. Parameters controlling the calculations
include the number of bands\,\cite{Shih:10,Friedrich:11,Klimes_Kresse:14}, the exchange-correlation
potential for the starting wave function\cite{Lany:10} as well as the use of
approximate models for the screening, like plasmon pole approximations \cite{Shih:10}.

Depending on the starting functional and the details of the
calculation, values between 2.1 and 3.6 eV are obtained for 
$G_0W_0$ \cite{Shishkin:07,Kang:14,Shih:10,Friedrich:11,Sarsari:13,Thygesen:2013,louie:2014,Klimes_Kresse:14}, 
between 2.54-3.6 for GW0 \cite{Sarsari:13,Kang:14,Friedrich:14pc,Thygesen:2013,Bruneval:2014} and 
between 3.2-4.3 for GW \cite{Usuda,Sarsari:13}. 

We start by validating the $GW_0$ method for bulk ZnO. For this purpose we consider a four-atom
wurzite ZnO unit cell and employ a 8$\times$8$\times$8 {\bf k} point
sampling with an energy cutoff of 500\,eV. The resulting band gaps and
energetic positions of the Zn-3$d$ states (with respect to the valence
band maximum set at zero) for several levels of GW calculations are a)
PBE: 0.8 and -5.1 eV, b) PBE0: 3.2 and -7.3\,eV, c) HSE06: 2.5 and
-7.1 eV, d) PBE+$GW$: 4.3 and -7.2\,eV, e) PBE+$GW_0$: 3.3 and -7.0
eV.

\begin{figure}[ht!]
\includegraphics[width=1\columnwidth,clip]{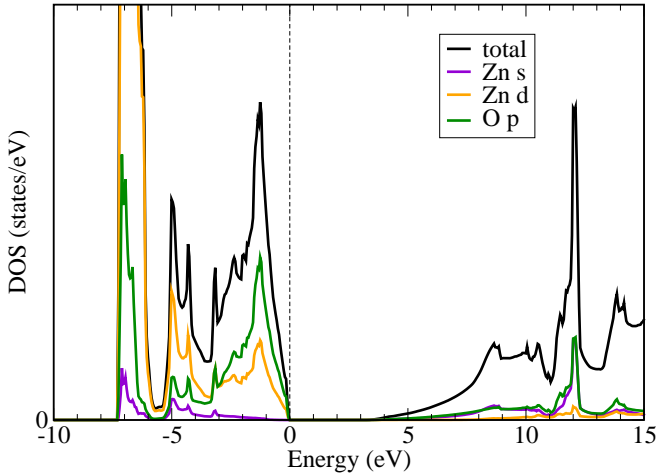} 
\caption{\label{fig:dos_bulk_gw0}(color online) Density of states of bulk ZnO calculated in the PBE+$GW_0$ approximation. The vertical line denotes the highest occupied state.}  
\end{figure}

For the $GW_{0}$ calculation, a cutoff of 200\,eV for the response
functions $\chi$, as well as 1024 bands have been employed. It has
been shown that a large number of bands is necessary to obtain
properly converged results in earlier one-shot $G_0W_0$ calculations
\cite{Shih:10,Friedrich:11,Klimes_Kresse:14}. We speculate that the
deviations to Ref.~\cite{Kang:14} are due to the rather low number of
bands in that publication.  The results are shown in
Fig.~\ref{fig:dos_bulk_gw0}. We find a band gap of 3.3\,eV in
reasonable agreement with the experimental value of 3.44\,eV
\cite{madelung:parameters}, as well as with other all-electron $GW_0$
calculations \cite{Friedrich:14pc,Klimes_Kresse:14}.  It should be
pointed out that recently norm conserving pseudopotentials were
introduced in VASP to improve $GW$ calculations
\cite{Klimes_Kresse:14}. However, as shown in Ref.~\cite{Klimes_Kresse:14}, we expect these to only lead to
minor modifications of the bandstructure of up to 0.2eV, not changing the main conclusions of this work.
It should be noted that that use of $GW$-based
methods has been able to predict the positions of lanthanide $f$-states
in NiO \cite{Aryasetiawan:97} and CeO$_2$ \cite{Jiang:09}.

The center of the Zn-$d$ orbitals is localted at -7\,eV, in agreement with previous
$GW_0$-calculations for ZnO \cite{Shishkin:07,Fuchs:07,Friedrich:14pc}.  We want
to point out, that a fully-selfconsistent $GW$ calculation results in
a much too high quasi-particle gap of 4.2\,eV.  Again this agrees
well with recent all-electron calculations  that find a quasi-particle gap of 4.3 \cite{Friedrich:14pc}.

\begin{figure}[ht!]
\includegraphics[width=0.8\columnwidth,clip]{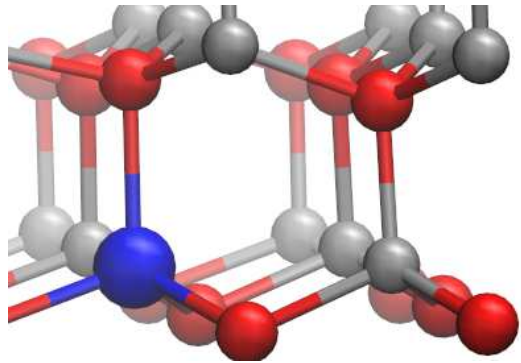} 
\caption{\label{fig:fig_eu}(color online) Atomic structure around the europium impurity calculated within GGA for substitutional europium in ZnO. Blue, red and grey spheres are Eu, O and Zn atoms, respectively.}
\end{figure}

From here on we show only results for GW$_0$ calculations.  It has
been shown that the position of defect states is not strongly
influenced by the number of empty bands and the response function
cutoff \cite{Sarsari:13}.  The relaxed geometry of Eu at a zinc
lattice position without the presence of intrinsic defects is shown in
Fig.~\ref{fig:fig_eu}.  The europium distance to nearest oxygen atoms
are 2.23{\AA} and 2.27{\AA} for in-plane and {\bf c}-direction,
respectively. These values are slightly larger than the Zn-O bond
lengths in pure ZnO. This means that if Eu is incorporated at a Zn
lattice position, it should not disturb the lattice significantly.
We would like to point out that in the supercell $GW_0$ calculations, the ZnO bandgap 
is slightly to large, around 3.6-3.9eV. We attribute this to the reduced $k$-point sampling in the supercell calculation.
We also several tests with sc$GW_0$ calculations, to also investigate the influence of a selfconsistent update of also the 
wave-functions. In these tests, we find that quasi-particle energies are only changed by up to 0.1-0.2eV between sc$GW_0$ and $GW_0$ calculations, 
supporting the use of the $GW_0$ approximation in the following.

\begin{figure}[h!] 
\includegraphics[width=1\columnwidth,clip]{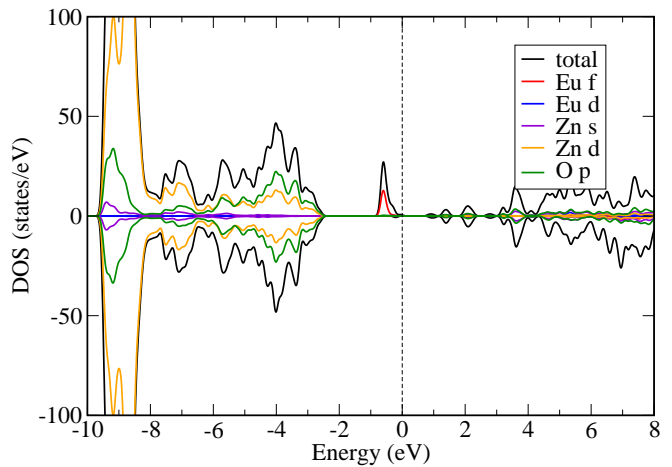} 
\caption{\label{fig:dos_eu}(color online) Total and projected density of states for Eu at Zn site in ZnO calculated within the PBE0+$GW_0$ approximation. The vertical dashed line denotes the highest occupied state. Positive (negative) values of the DOS denote spin up (down).}
\end{figure}

In Fig.~\ref{fig:dos_eu} the $GW_0$ electronic density-of-states of
substitutionally incorporated Eu in this geometry is shown. The Eu-$f$ states lie within
the band gap, around 2\,eV above the VBM (valence band maximum),
hybridize weakly with the Zn-$s$ and O-$p$ states. As expected for
substitutional Eu at a Zn lattice position, Eu has a formal charge
closer to 2+. The $f$ spin-up orbitals are fully occupied, giving a
total magnetic moment of 7$\mu_{\rm B}$. Furthemore, the
Eu $d$-states lie deep in the conduction band of ZnO, $\approx$2.5\,eV
above the CBM, leading to an energetic difference to the Eu $f$-states
of over 3eV. 
Therefore, our results explain why the $4f^7-4f^6/5d^1$ optical transition at 530\,nm ($\approx$ 2.3 eV)
that is often observed additionally to the intra-f transitions for Eu incorporation into other host materials
can not be found experimentally in Eu doped ZnO \cite{Petersen:10,Ronning:10}.

\begin{figure}[!h]
\includegraphics[width=0.8\columnwidth,clip]{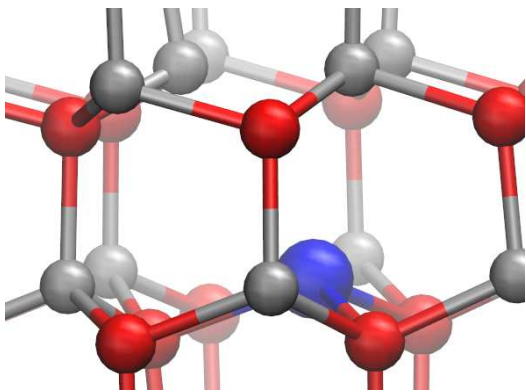}
\caption{\label{fig:fig_euov}(color online) Atomic structure around the ${\rm Eu-V_O}$ complex calculated within GGA.  Blue, red and gray spheres are Eu, O and Zn atoms, respectively.}
\end{figure}

A possible way to modify the oxidation state of Eu in the ZnO lattice
is therefore a change of its environment. It is known that during ion
implantation intrinsic defects are likely to form
\cite{Ronning:10,Wang:11}.  We therefore investigated Eu doped ZnO in
the presence of nearby oxygen and zinc vacancies as well as zinc and
oxygen interstitials.
In the presence of a neutral oxygen vacancy the
Eu-O distances for neighboring oxygen atoms remain almost unchanged,
compared to the case without vacancy, and increase slightly to
2.24\AA. The geometry is shown in Fig.~\ref{fig:fig_euov}.  The Eu-Zn
distance is 3.27\AA and relaxation of farther neighbors is
insignificant.

\begin{figure}[!h]
\includegraphics[width=1\columnwidth,clip]{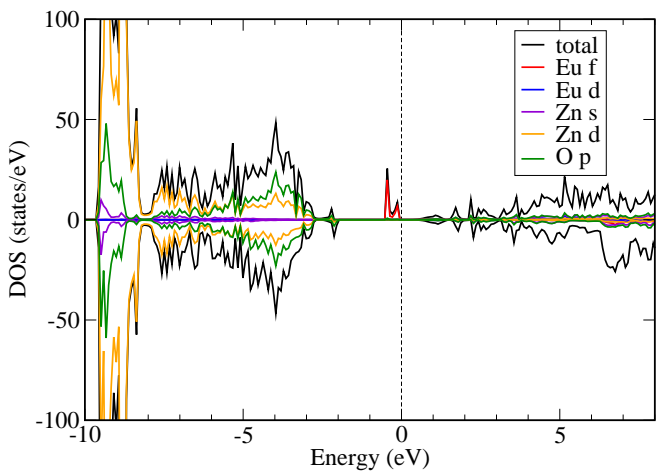}
\caption{\label{fig:dos_euov}(color online) Total and projected density of states for Eu incorporated at Zn position in ZnO next to a oxygen vacancy calculated within the PBE0+$GW_0$ approximation. The vertical line denotes the highest occupied state. Positive (negative) values of the DOS denote spin up (down).}
\end{figure}

The electronic structure for this system is shown in
Fig.~\ref{fig:dos_euov}. The states due to the oxygen vacancy lie
1\,eV above the VBM. This is in good agreement with calculations for
pure ZnO (see e.g. Refs.~\cite{Janotti:09,Oba:10,Oba:11}). The Eu $f$-states
are located now at 2.5\,eV above the VBM.  Moreover, we observe a
small splitting of the Eu-$f$ states as well as a small hybridization
with the Eu-$d$ and O-$p$ states.  The splitting is caused by a shift
of one of the seven Eu-$f$ states that we assume to be induced by
changes in the local environment that affect the interaction matrix elements and therefore the charge distribution 
around the impurity.

Our GW$_0$ results are in
strong contrast to the findings of Ref.~\cite{Assadi:11}, where the
Eu-$f$ states are energetically located directly above the V$_O$
states. We attribute this discrepancy to the choice of the GGA
functional in Ref.~\cite{Assadi:11}, predicting a much too narrow band
gap.  On the other hand, the presence of the $\text{V}_\text{O}$ does
not significantly change the formal charge around the Eu atom.
Moreover, the positions of the Eu-$f$ spin down and Eu $d$-states
remain practically unchanged. Again this defects cannot explain the
experimental observed $f-f$ transition in ZnO.

\begin{figure}[t]
\includegraphics[width=0.8\columnwidth,clip]{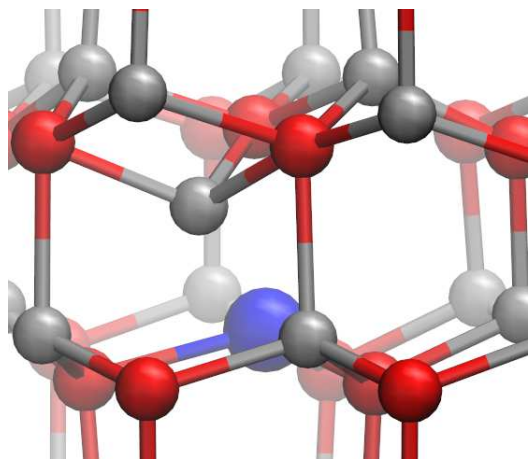}
\caption{\label{fig:fig_euznint}(color online) Atomic structure around the ${\rm Eu-Zn_i}$ complex calculated within GGA. Blue, red and gray spheres are Eu, O and Zn atoms, respectively.}
\end{figure}

The relaxed geometry of the Eu+Zn$_i$ defect is shown in
Fig.~\ref{fig:fig_euznint}. The bond lengths between the Eu atom and the
nearest oxygen atoms are 2.19{\AA}, 2.19{\AA} and 2.24{\AA} in the
in-plane direction, and 2.41{\AA} in the ${\bf c}$-direction. This is
because Eu atom relaxes away from the Zn atom. The Eu-Zn distance is
2.72{\AA}, while the distances to Zn second neighbors are 3.29{\AA}.

\begin{figure}[!h]
\includegraphics[width=1\columnwidth,clip]{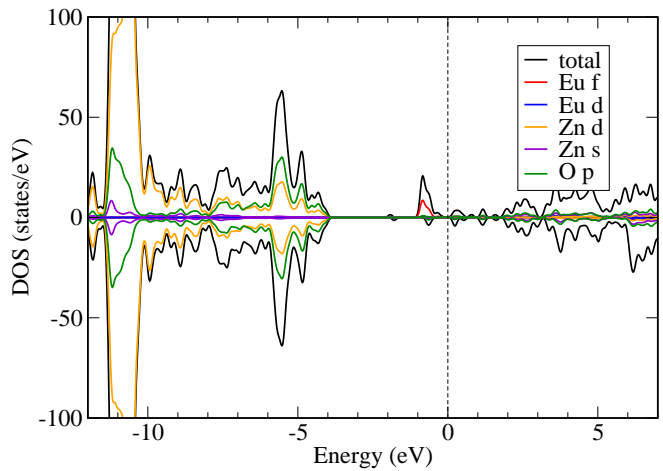}
\caption{\label{fig:dos_euznint}(color online) Density of states of Eu doped ZnO in the presence of a nearby zinc interstitial calculated within the PBE0+GW0 approximation. The vertical line denotes the highest occupied state. Positive (negative) values are the spin up (down) components.}
\end{figure}

The electronic structure of this system is shown in
Fig.~\ref{fig:dos_euznint}.  The location of the Eu $f$-states lies
close to the conduction band minimum (CBM), around 2.9\,eV above the
VBM. We can infer a weak hybridization with both Zn-$s$ and Zn-$d$
states of the Zn interstitial atom. Similarly, a very small
hybridization with the O-$p$ states is found. Here, the formal charge
of the Eu atom remains close to 2+ with the spin-up $f$-states
occupied. 

\begin{figure}[!h]
\includegraphics[width=0.8\columnwidth,clip]{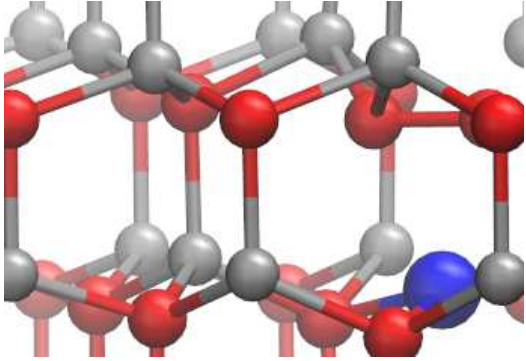}
\caption{\label{fig:fig_eu_oint_struc}(color online) Atomic structure around the ${\rm Eu-O_{int}^{split}}$  complex calculated within GGA. Blue, red and gray spheres are Eu, O and Zn atoms, respectively.}
\end{figure}

\begin{figure}[!h]
\includegraphics[width=1\columnwidth,clip]{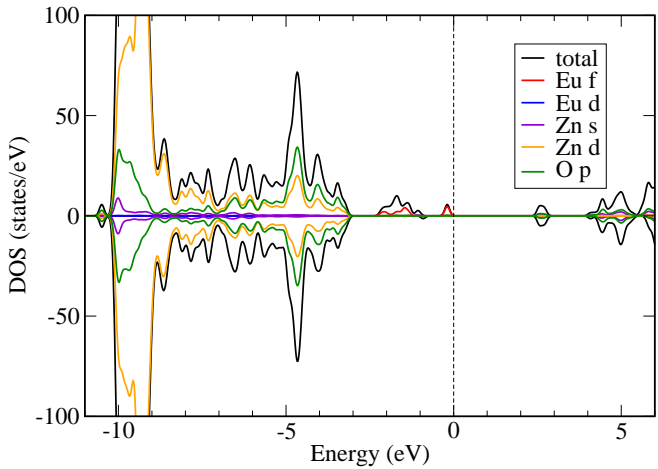}
\caption{\label{fig:dos_eu_oint_split}(color online) Density of states of Eu doped ZnO in the presence of a nearby split oxygen interstitial calculated within the PBE0+GW0 approximation. The vertical line denotes the highest occupied state. Positive (negative) values are the spin up (down) components.}
\end{figure}

For the oxygen interstitial plus Eu defect, we have considered two
geometries.  The first consists of an Eu atom at substitutional Zn
site next to an oxygen interstitial in dumbbell configuration \cite{Erhart:05}. 
This configuration is shown in Fig.~\ref{fig:fig_eu_oint_struc}. 
In this case the lattice ralaxations are more pronounced,
The Eu atom is shifted away from the interstitial complex, decreasing the Eu-O bond lengths to 
2.21\AA. The Eu-Zn distance is also slightly increased to 3.29\AA and the distances to the oxygen atoms of the interstitial 
are 2.31\AA and 2.33\AA, respctively.

The electronic structure is presented in Fig.~\ref{fig:dos_eu_oint_split}.
We observe a splitting of the Eu-$f$ states that we attribute to a combination of the changed symmetry in the local environment.
This affects both the interaction matrix elements 
as well as their screening, so that different Eu states are influences differently.
While 6 of the Eu-$f$ are located closer to the 
VBM, the remaining occupied Eu-$f$ state lies about 1 eV higher in energy.
We find contributions to the localized states from both ${\rm O_{int-p}}$
and ${\rm O_{host}}$ states. However, also in this case, the Eu atom possesses a formal charge of 2+
with all Eu-$f$ spin-up states fully occupied. The oxygen interstitial atom possesses a magetic moment of 0$\mu_B$,
a further confirmation, that no electron is transfered from the Eu atom to the oxygen site.

\begin{figure}[!h]
\includegraphics[width=0.8\columnwidth,clip]{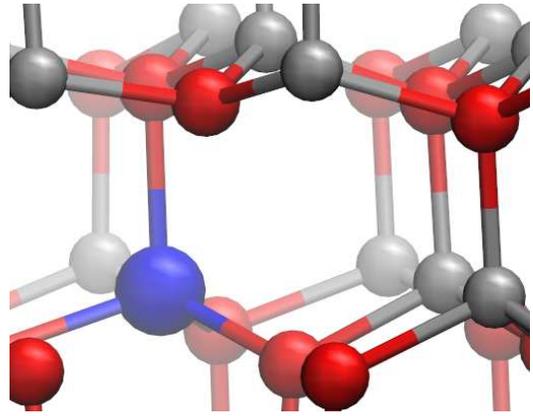}
\caption{\label{fig:fig_euzv}(color online) Atomic structure around the ${\rm Eu-V_{Zn}}$ complex within GGA. Blue, red and gray spheres are Eu, O and Zn atoms, respectively.}
\end{figure}

\begin{figure}[h]
\includegraphics[width=1\columnwidth,clip]{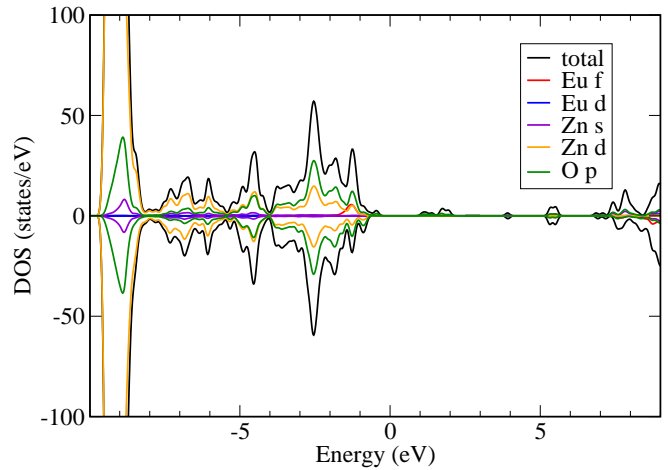}
\caption{\label{fig:dos_euzv}(color online) Density of states of Eu doped ZnO in the presence of a nearby zinc vacancy calculated within the PBE0+GW0 approximation. The vertical line denotes the highest occupied state. Positive (negative) values are the spin up (down) components.}
\end{figure}

The geometry of Eu doped ZnO in the presence of a neutral Zn vacancy
(V$_{\rm Zn}$) is shown in Fig.~\ref{fig:fig_euzv}. The Eu atom shifts
towards the vacancy, increasing the in-plane Zn-Eu distance to
3.49\AA{} and 3.40\AA{} in comparison to 3.3\AA{} that would
correspond to Eu exactly on Zn lattice position.  Consequently, the
in-plane Eu-O distances also change, asymmetrically, to 2.09\AA{},
2.19\AA{} and 2.24\AA{}, respectively.  Along the {\bf c}-direction
the Eu-O distance is 2.22\AA{}. The corresponding electronic structure
is shown in Fig.~\ref{fig:dos_euzv}.  We find no occupied Eu-$f$
states in the band gap.  These states lie within the VB, close to the
O-$p$ states and hybridize both with with O-$p$ and Zn-$d$ states. The
local projected moment on the V$_{\rm Zn}$ site is 1$\mu_B$, which is
aligned anti-parallel to the magnetic moment on the Eu which is
6$\mu_{\rm B}$.   However the fact that the Eu-$f$ states are
energetically located inside the VB and the unoccupied Eu-$f$ state is
energetically located at 2 eV above the VBM, does not fit with
experimentally observed transition.

\begin{figure}[!h]
\includegraphics[width=0.8\columnwidth,clip]{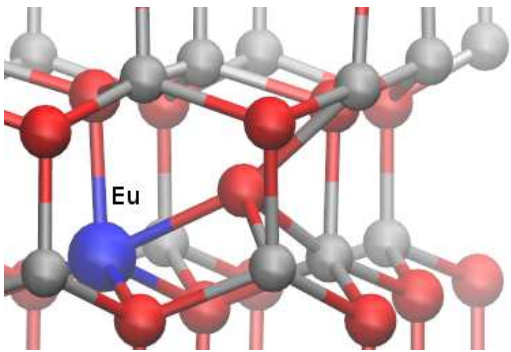}
\caption{\label{fig:fig_eu_oint_oct_struc}(color online) Atomic structure around the Eu+O$_\text{int}^\text{oct}$ complex within GGA. Blue, red and gray spheres are Eu, O and Zn atoms, respectively.}
\end{figure}

The other defect complex including an oxygen interstitial consists of
an oxygen at a octahedral interstitial site next to an Eu at Zn
lattice position.  The corresponding atomic structure is shown in
Fig.~\ref{fig:fig_eu_oint_oct_struc}.  We find much more pronounced
lattice relaxations in this case, also involving second nearest
neighbors. The Eu-Zn distance is 3.2\AA{}, while the Eu-O bond lengths
are 2.2\AA{} to the oxygen atoms in the basal plane and 2.25\AA{} to
the oxygen along the $c$-direction.  The distance between the Eu atom
and the oxygen interstitial is 2.24\AA{}.  This can be explained by
looking at the coordination Eu adopts. The Eu-O distance to the
nearest neighbors is very similar to Eu$_2$O$_3$, which possess a
coordination number of $6\pm 1$ and a radial distance of 2.33
$\pm  0.015$ {\AA} according to Ref\cite{Fryxell:2004}.

\begin{figure}[!h]
\includegraphics[width=1\columnwidth,clip]{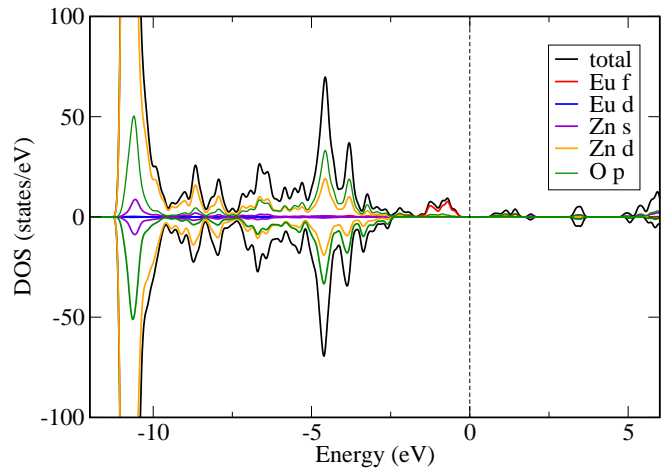}
\caption{\label{fig:dos_eu_oint_dos}(color online) Density of states of Eu doped ZnO in the presence of a nearby octahedral oxygen interstitial calculated within the PBE0+GW0 approximation. The vertical dashed line denotes the highest occupied state. Positive (negative) values are the spin up (down) components.}
\end{figure}

The corresponding electronic structure is shown in
Fig.~\ref{fig:dos_eu_oint_dos}. We find the occupied Eu $f$-states are
located in the band gap, with the Eu-$f$ orbitals being occupied with
six electrons. The occupied and unoccupied states are located at
-0.7\ eV and 1.3\ eV, respectively.  We find a small hybridization
with O-$p_\text{i(oct)}$ and O-$p_\text{(host)}$. The total magnetic moment of
the complex is 5$\mu_{\rm B}$, determined by a magnetic moment of
6$\mu_{\rm B}$ spin up at the europium aligned anti-parallel to the
magnetic moment of 1$\mu_{\rm B}$ at the oxygen interstitial atom.
Therefore, we suggest that the intra-4$f$ luminescence of the ZnO:Eu
samples is most likely due to Eu$_\text{Zn}$+O$^\text{oct}_i$
complexes.  We have also explored the mechanism for optical activation
of Eu in ZnO reported in Ref.~\cite{Akazawa:13}, where substitutional
hydrogen at Zn lattice position next to an europium atom is
suggested. However, this configuration does not change the formal
charge of Eu.  

In order to summarize the results discussed above we
show the position of the defect levels with respect to the band edges of
ZnO. The results are shown in Table \ref{tab:edge}. We show the relative position of Eu-f states for the complexes with respect to the top of the valence band of ZnO. The position of the Eu-f states were taken approximately in the middle of the $f$-bands, since we cannot describe correctly with DFT the multiplets of the f-orbitals.

\begin{table}[ht!]
\begin{center}
\caption{\label{tab:edge} Relative position of Eu-f states for the investigated complexes with respect to the top of the valence band of ZnO. The position of the Eu-f states were taken approximately in the middle of the f-bands. All values are given in eV.}
\begin{tabular*}{0.5\textwidth}{@{\extracolsep{\fill}}lc}
\hline
\hline
Complex  & 	{${\rm E_{VB}-E_f}$}  	\\
\hline
\hline
${\rm Eu_{Zn}}$ 	&  2.8   \\
${\rm Eu_{Zn}+O{i(split)}}$ & 1.5/3.2   \\
${\rm Eu_{Zn}+O{i(oct)}}$ & 2.5    \\
${\rm Eu_{Zn}+V_{O}}$ &  2.5/3.2 \\
${\rm Eu_{Zn}+V_{Zn}}$ & 1.0  \\
${\rm Eu_{Zn}+Zn_{int}}$ & 3.8    \\
\hline
\hline
\end{tabular*}
\end{center}
\end{table}

\section{Conclusions}

In conclusion, we have investigated Eu-doped ZnO using DFT and the
many-body GW technique within the GW$_0$ approximation.  We find that
the position and formal charge of the Eu-$f$ states is strongly
dependent on the environment around the Eu atom. We conclude that the
optical activity of Eu in ZnO is most likely due to Eu+O$^\text{oct}_\text{i}$
defect complexes, possibly either in a neutral or -1 charge state or in the presence of both.  
Finally, we believe our results can open the pathway
for a better understanding of these complexes in zinc oxide.

\begin{acknowledgments}
We acknowledge fruitful discussions with R. R\"oder, C. Ronning and H. Chacham and
financial support from the Deutsche Forschungsgemeinschaft under the
programm FOR1616. A. L. da Rosa would like to thank financial support
from CNPq under the program ``Science without Borders''. We also thank
HLRN (Hannover/Berlin) for computational resources.
\end{acknowledgments}


\end{document}